\documentclass[aps,pre,twocolumn,floats]{revtex4}
\usepackage{epsfig}
\usepackage{bm}

\begin{document}

\newcommand{\hide}[1]{}
\newcommand{\tbox}[1]{\mbox{\tiny #1}}
\newcommand{\half}{\mbox{\small $\frac{1}{2}$}}
\newcommand{\const}{\mbox{const}}
\newcommand{\ointt}{\int\!\!\!\!\int\!\!\!\!\!\circ\ }
\newcommand{\intt}{\int\!\!\!\!\int }
\newcommand{\ar}{\mathsf r}
\newcommand{\im}{\mbox{Im}}
\newcommand{\re}{\mbox{Re}}
\newcommand{\mbf}[1]{#1}
\newcommand{\bmsf}[1]{\bm{\mathsf{#1}}} 
\newcommand{\fullG}{G}
\newcommand{\symG}{\eta}
\newcommand{\sinc}{\mbox{sinc}}
\newcommand{\trc}{\mbox{trace}}
\newcommand{\eexp}{\mbox{e}^}
\newcommand{\bra}{\left\langle}
\newcommand{\ket}{\right\rangle}


\title{
Quantum pumping in closed systems, adiabatic transport, and the Kubo formula
}

\author{Doron Cohen}

\affiliation{
Department of Physics, Ben-Gurion University, Beer-Sheva 84105, Israel
}


\begin{abstract}
Quantum pumping in closed systems is considered.
We explain that the Kubo formula contains all
the physically relevant ingredients for the calculation
of the pumped charge ($Q$) within the framework
of linear response theory. The relation to
the common formulations of adiabatic transport
and ``geometric magnetism" is clarified.
We distinguish between adiabatic and dissipative
contributions to~$Q$.
On the one hand we observe that adiabatic pumping does
not have to be quantized. On the other hand we
define circumstances in which quantized adiabatic pumping
holds as an approximation. The deviation from
exact quantization is related to the Thouless conductance.
As an application we discuss the following examples:
classical dissipative pumping by conductance control,
classical adiabatic (non dissipative) pumping by translation,
and quantum pumping in the double barrier model.
In the latter context we analyze a 3~site lattice Hamiltonian,
which represents the simplest pumping device.
We remark on the connection with the popular $S$ matrix formalism
which has been used to calculate pumping in open systems.
\end{abstract}

\maketitle

\section{Introduction}

Linear response theory (LRT) \cite{landau,imry,datta}
is the leading formalism to
deal with driven systems. Such systems are described
by a Hamiltonian ${\cal H}(\bm{x})$ where $\bm{x}(t)$ is a set
of time dependent classical parameters ("fields").
The Kubo formula is the corner stone of LRT.
It allows the calculation of the response coefficients,
and in particular the {\em conductance matrix} ($\bmsf{G}$)
of the system. If we know $\bmsf{G}$, we can calculate
the charge ($Q$) which is transported through the
system during one cycle of a periodic driving.
This is called {\em pumping}.

Pumping of charge in mesoscopic \cite{marcus_rev}
and molecular size devices is regarded as a major
issue in the realization of future quantum circuits
or quantum gates, possibly for the purpose of
quantum computing.

\subsection{Model system}

In order to explain the motivation for the present work,
and its relation to the published literature,
we have to give a better definition of the problem.
For presentation purpose we focus on
a model system with a ring geometry (Fig.1).
The shape of the ring is controlled
by some parameters $x_1$ and $x_2$.
These parameters can be gate voltages that determine
the location of some boundaries, or the height of some barriers.
The third parameter is the flux through the ring:
\begin{eqnarray}
x_3 \ = \ \Phi \ \equiv \ (\hbar/e) \phi
\end{eqnarray}
We shall use units such that the
elementary charge is \mbox{$e=1$}.
Note that the Hamiltonian ${\cal H}(x_1(t),x_2(t),x_3(t))$
has gauge invariance for $\phi \mapsto \phi+2\pi$.
Another system with a ring topology is presents in Fig.~1b,
and its abstraction is represented in Fig.~2c.
The ``dot" can be represented by an $S$ matrix
that depends on $x_1$ and $x_2$.
In Fig.~1d also the flux $x_3$ is regarded as a parameter of the dot.
If we cut the wire in Fig.1d we get the open two lead geometry
of Fig.1e. Finally we can put many such units in series (no flux),
hence getting the periodic system of Fig.1f.
In the latter case the Hamiltonian is invariant for
unit translations, and therefore the quasi momentum $\phi$
is a constant of motion. It follows that the mathematical
treatment of  a driven periodic structure reduces to
an analysis of a driven ring system with flux.

\subsection{Classification of pumps}

{\em ``Pumping"} means that net charge (or maybe better
to say ``net integrated probability current")  is transported
through the ring per cycle of a periodic driving.
Using the common jargon of electrical engineering
this can be described as AC-DC conversion.
We distinguish between \\
\begin{minipage}{\hsize}
\vspace*{0.1cm}
\begin{itemize}
\setlength{\itemsep}{0cm}
\item
pumping in open systems (such as in Fig.~1e).
\item
pumping in closed systems (such as in Fig.~1d)
\item
pumping in periodic systems (such as in Fig.~1f)
\end{itemize}
\vspace*{0.0cm}
\end{minipage}
For a reason that was explained at the end of the previous subsection
we regard the last category \cite{thouless} as mathematically equivalent
to the second category. We also regard the first
category \cite{BPT,aleiner,barriers,ora}
as a special (subtle) limit of the second category:
in a follow up paper \cite{pmo} we demonstrate
that in the limit of open geometry the Kubo formula
reduces to the $S$-matrix formula
of B\"{u}ttiker Pr\'{e}tre and Thomas \cite{BPT}.

There are works in the literature regarding
``rectification" and ``ratchets" \cite{ratch}.
These can be regarded as studies of
pumping in periodic systems 
with the connotation of having  {\em damped}
non-Hamiltonian dynamics. These
type of systems are beyond the scope
of the present Paper.
There is also a recent interest in Hamiltonian Ratchets\cite{ratchets},
which is again a synonym for pumping in periodic systems,
but with the connotation of having a {\em non-linear}
pumping mechanism.
We are going to clarify what are the conditions for having
a {\em linear} pumping mechanism. Only in case of linear pumping mechanism
the Kubo formula can be used, which should be distinguished
from the non-linear mechanism of Ref.\cite{ratchets}.

\subsection{Objectives}

The purpose of this Paper is to explain and demonstrate
that the Kubo formula contains all
the physically relevant ingredients for
the calculation of the charge ($Q$) which is pumped
during one cycle of a periodic driving.
In the limit of a very slow time variation (small $\dot{x}$),
the emerging picture coincides with
the adiabatic picture of Refs.\cite{berry,thouless,avron,robbins}.
In this limit the response of the system
is commonly described as a non-dissipative
``geometric magnetism"  \cite{robbins} effect,
or as adiabatic transport.

A major objective of this Paper is to bridge between
the adiabatic picture  and the more general LRT / Kubo picture,
and to explain how {\em dissipation} emerges
in the quantum mechanical treatment.
For one-parameter driving a unifying picture that bridges between
the quantum mechanical adiabatic picture and LRT has been presented
in \cite{vrn,dsp,crs,frc}. A previous attempt \cite{wilk} had ended
in some confusion regarding the identification of the linear
response regime, while \cite{robbins} had avoided the analysis of
the mechanism that leads to dissipation in the quantum mechanical case.

The presented (Kubo based) formulation of the pumping problem
has few advantages:
It is not restricted  to the adiabatic regime;
It allows a clear distinction between {\em dissipative} and {\em adiabatic} contributions to the pumping;
The classical limit is manifest in the formulation;
It gives a level by level understanding of the pumping process;
It allows the consideration of any type of occupation (not necessarily Fermi occupation);
It allows future incorporation of external environmental influences such as that of noise;
It regards the voltage over the pump as electro motive force,
rather than adopting the conceptually more complicated view \cite{ora}
of having a chemical potential difference.

Of particular interest is the possibility to realize
a pumping cycle that transfers {\em exactly} one unit of charge
per cycle. In open systems \cite{aleiner,barriers}
this ``quantization" holds only approximately,
and  it has been argued \cite{aleiner} that the deviation from
exact quantization is due to the dissipative effect.
Furthermore it has been claimed \cite{aleiner}
that exact quantization would hold in the strict adiabatic limit,
if the system were {\em closed}.
In this Paper we would like to show that the correct picture is quite different.
We shall demonstrate  that the deviation from exact quantization is in fact
of adiabatic nature. This deviation is related
to the so-called ``Thouless conductance" of the device.

\subsection{Examples}

We give several examples for the application
of the Kubo formula to the calculation of
the pumped charge~$Q$: \\
\begin{minipage}{\hsize}
\vspace*{0.1cm}
\begin{itemize}
\setlength{\itemsep}{0cm}
\item
classical dissipative pumping
\item
classical adiabatic pumping (by translation)
\item
quantum pumping in the double barrier model
\end{itemize}
\vspace*{0.0cm}
\end{minipage}
The last example is the main one.
In the context of open geometry it is known
as ``pumping around a resonance" \cite{barriers}.
We explain that this is in fact
an diabatic transfer scheme,
and we analyze a particular version
of this model which is represented
by a 3~site lattice Hamiltonian.
This is definitely the simplest pump circuit
possible, and we believe that it can be realized
as a molecular size device.
It also can be regarded as an approximation
for the closed geometry version of the
two delta potential pump \cite{barriers}.

\subsection{Outline}

In Section~2  we define the main object of the study,
which is the conductance matrix $\bmsf{G}$ of Eq.(\ref{e_8}).
The conductance matrix can be written as the sum
of a symmetric ($\bm{\eta}$) and an anti-symmetric ($\bmsf{B}$)
matrices, which are later identified as the dissipative and
the adiabatic contributions respectively.

In the first part of the paper (Sections~2-8)
we analyze the adiabatic equation (Section~3),
and illuminate the distinction
between its zero order solution (Section~4),
its stationary first order solution (Section~5),
and its non-stationary solution (Section~6).
The outcome of the analysis in Section~5 is Eq.(\ref{e_30})
for the conductance matrix $\bmsf{G}$.
This expression is purely adiabatic, and does
not give any dissipation. In order to get dissipation
we have to look for a non-stationary solution.

The standard textbook derivation of the Kubo formula (Eq.(\ref{e_33}))
for the conductance matrix $\bmsf{G}$ {\em implicitly} assumes
a non-stationary solution. We show how to get from it
Eq.(\ref{e_34}) for $\bm{\eta}$ and Eq.(\ref{e_35}) for $\bmsf{B}$.
The latter is shown to be identical with the adiabatic result (Eq.(\ref{e_30})).
In Section~7 we further simplify the expression for $\bm{\eta}$
leading to the fluctuation-dissipation relation (Eq.(\ref{e43})).

The disadvantages of the standard textbook derivation
of Kubo formula make it is essential to introduce a different
route toward  Eq.(\ref{e43})  for $\bm{\eta}$.
This route, which is discusssed in Section~8,
{\em explicitly} distinguishes the dissipative effect from the adiabatic effect,
and allows to determine the conditions for the validity
of either the adiabatic picture or LRT.
In particular it is explained that LRT is based,
as strange as it sounds, on perturbation theory to infinite order.

In Section~9 we clarify the general scheme of
the pumping calculation (Eq.(\ref{e6})).
Section~10 and Section~11 give two simple classical examples.
In Section~12 we turn to discuss quantum pumping,
where the cycle is around a chain of degeneracies.
The general discussion is followed by presentation
of the double barrier model (Section~13).
In order to get a quantitative estimate for the pumped charge
we consider a 3~site lattice Hamiltonian (Section~14).

The summary (Section~15) gives some larger perspective on the subject,
pointing out the relation to the $S$-matrix formalism,
and to the Born-Oppenheimer picture.
In the appendices we give some more details regarding the
derivations, so as to have a self-contained presentation.

\section{The conductance matrix}

Consider the Hamiltonian ${\cal H}(\bm{x}(t))$, where $\bm{x}(t)$
is a set of time dependent parameters ("fields").
For presentation, as well as for practical reasons,
we assume later a set of three time dependent parameters
\mbox{$\bm{x}(t)=(x_1(t),x_2(t),x_3(t))$}.
We define generalized forces in the conventional way as
\begin{eqnarray}
F^k \ \ = \ \ -\frac{\partial {\cal H}}{\partial x_k}
\end{eqnarray}
Note that if $x_1$ is the location of a
wall element, then $F^1$ is the force in the
Newtonian sense. If $x_2$ is an electric field,
then $F^2$ is the polarization.
If $x_3$ is the magnetic field or the flux through a ring,
then $F^3$ is the magnetization or the current through the ring.

In linear response theory (LRT) the response of the system
is described by a causal response kernel, namely
\begin{eqnarray} \label{e_3}
\langle F^k \rangle_t \ = \ \sum_j \int_{-\infty}^{\infty} \alpha^{kj}(t-t') \ x_j(t')dt'
\end{eqnarray}
where $\alpha^{kj}(\tau)=0$ for $\tau<0$.
The Fourier transform of  $\alpha^{kj}(\tau)$
is the generalized susceptibility $\chi^{kj}(\omega)$.
The conductance matrix is defined as:
\begin{eqnarray}
\fullG^{kj} \ = \ \lim_{\omega\rightarrow 0}
\frac{\im[\chi^{kj}(\omega)]}{\omega}
\ = \ \int_0^{\infty} \alpha^{kj}(\tau) \tau d\tau
\end{eqnarray}
Consequently, as explained further in Appendix~B,
the response in the ``DC limit" ($\omega\rightarrow0$)
can be written as
\begin{eqnarray} \label{e_8}
\langle F^k \rangle =
-\sum_{j} \fullG^{kj} \ \dot{x}_j
\end{eqnarray}
As an example for the applicability of this formula
note the following standard examples for one parameter driving:
Let $x=$ wall or piston displacement, then $\dot{x}=$ wall or piston velocity,
$G=$ friction coefficient, and $F=-G \dot{x}$ is the friction force.
Another standard example is $x=$ magnetic flux,
$-\dot{x}=$ electro motive force, $G=$ electrical conductance,
and hence $F=-G \dot{x}$ is Ohm law.

It is convenient to write the conductance matrix
as  $\fullG^{kj} \equiv \ \symG^{kj} + \mbf{B}^{kj}$,
where $\symG^{kj}=\symG^{jk}$ is the symmetric part
of the conductance matrix,
while $\mbf{B}^{kj}=-\mbf{B}^{jk}$ is the antisymmetric part.
In case of having three parameters we can arrange the
elements of the antisymmetric part
as a vector \mbox{$\vec{\bm{B}}=(B^{23},B^{31},B^{12})$}.
Consequently Eq.(\ref{e_8}) can be written in abstract notation as
\begin{eqnarray}
\langle \bm{F} \rangle \ = \ -\bm{\symG} \cdot \dot{\bm{x}} \ - \ \bmsf{B}\wedge \dot{\bm{x}}
\end{eqnarray}
where the dot product should be interpreted as matrix-vector
multiplication, which involves summation over the index~$j$.
The wedge-product also can be regarded as a matrix-vector
multiplication.  It reduces to the more familiar cross-product
in case that we consider $3$ parameters.
The dissipation, which is defined as the rate in which energy
is absorbed into the system, is given by
\begin{eqnarray}
\dot{{\cal W}}  = \frac{d}{dt}\langle {\cal H} \rangle
\ = \  -\langle \bm{F} \rangle \cdot \dot{\bm{x}} \ = \
\sum_{kj} \symG^{kj} \ \dot{x}_k\dot{x}_j
\end{eqnarray}
Only the symmetric part contributes
to the the dissipation. The contribution
of the antisymmetric part is identically zero.

\section{The adiabatic equation}

The adiabatic equation is conventionally obtained from the
Schrodinger equation by expanding the wavefunction
in the $x$-dependent adiabatic basis:
\begin{eqnarray}
\frac{d}{dt}|\psi\rangle \ \ &=&
\  -\frac{i}{\hbar} {\cal H}(\bm{x}(t)) \ |\psi\rangle
\\
|\psi\rangle \ \ &=& \ \ \sum_n a_n(t) \ |n(\bm{x}(t))\rangle
\\
\frac{da_n}{dt} &=& -\frac{i}{\hbar} E_n a_n
+\frac{i}{\hbar} \sum_m \sum_j
\dot{x}_j \mbf{A}^{j}_{nm} a_m
\end{eqnarray}
where following \cite{berry}  we define
\begin{eqnarray}
\mbf{A}^{j}_{nm}(\bm{x}) =
i\hbar \left\langle n(\bm{x}) \Big|
\frac{\partial}{\partial x_j} m(\bm{x})\right\rangle
\end{eqnarray}
Differentiation by parts of
$\partial_j \langle n(\bm{x})|m(\bm{x}) \rangle  = 0$
leads to the conclusion that $\mbf{A}^{j}_{nm}$
is a hermitian matrix.
Note that the effect of gauge transformation is
\begin{eqnarray}
|n(\bm{x})\rangle \ & \ \mapsto \ & \ \eexp{-i\frac{\Lambda_n(\bm{x})}{\hbar}} \ |n(\bm{x})\rangle \\
\mbf{A}^{j}_{nm} \ & \ \mapsto \ & \
\eexp{i\frac{\Lambda_n{-}\Lambda_m}{\hbar}} \mbf{A}^{j}_{nm}
+ (\partial_j \Lambda_n)\delta_{nm}
\end{eqnarray}
Note that the diagonal elements $\mbf{A}^{j}_{n} \equiv\mbf{A}^{j}_{nn}$
are real, and transform as
$\mbf{A}^{j}_{n} \mapsto \mbf{A}^{j}_{n} + \partial_j \Lambda_n$.

Associated with $\bm{A}_n(\bm{x})$
is the gauge invariant 2-form, which is defined as:
\begin{eqnarray}
\mbf{B}^{ij}_{n} &=&
\partial_i \mbf{A}^{j}_{n} - \partial_j \mbf{A}^{i}_{n} \\
&=& -2\hbar \im\langle\partial_i n|\partial_j n\rangle  \\
&=& -\frac{2}{\hbar}\im\sum_{m}
\mbf{A}^{i}_{nm}\mbf{A}^{j}_{mn}
\end{eqnarray}
This can be written in an abstract notation as $\bmsf{B}=\nabla\wedge\bm{A}$.
Using standard manipulations, namely via
differentiation by parts of
$\partial_j \langle n(\bm{x})|{\cal H}|m(\bm{x}) \rangle  = 0$,
we get for $n\ne m$ the expressions:
\begin{eqnarray}
\mbf{A}^{j}_{nm}(\bm{x}) = \frac{i\hbar}{E_m{-}E_n}
\left\langle n \left|\frac{\partial {\cal H}}{\partial x_j}\right|m\right\rangle
\equiv-\frac{i\hbar F^j_{nm}}{E_m{-}E_n}
\end{eqnarray}
and hence
\begin{eqnarray} \label{e29}
\mbf{B}^{ij}_{n} = 2\hbar \sum_{m(\ne n)}
\frac{\im\left[
F^i_{nm}F^j_{mn}\right]}
{(E_m-E_n)^2}
\end{eqnarray}

\section{The strictly adiabatic solution, and the Berry phase}

We define the perturbation matrix as
\begin{eqnarray} \label{e_23}
\mbf{W}_{nm}=-\sum_j\dot{x}_j\mbf{A}^j_{nm}
\ \ \ \ \ \ \ \ \mbox{for $n\ne m$}
\end{eqnarray}
and $\mbf{W}^{j}_{nm}=0$ for $n=m$.
Then the adiabatic equation can be re-written
as follows:
\begin{eqnarray} \label{e_24}
\frac{da_n}{dt} = -\frac{i}{\hbar} (E_n{-}\dot{x}A_n) a_n
-\frac{i}{\hbar} \sum_m
\mbf{W}_{nm} a_m
\end{eqnarray}
If we neglect the perturbation $W$,
then we get the strict adiabatic solution:
\begin{eqnarray}
\eexp{
-\frac{i}{\hbar} \left(
\int_0^t E_n(\bm{x}(t'))dt'
-\int_{\bm{x}(0)}^{\bm{x}(t)} \mbf{A}_n(\bm{x})\cdot d\bm{x}
\right)
}
\ |n(\bm{x}(t))\rangle
\end{eqnarray}
Due to $\bm{A}_n(\bm{x})$, we have the so called geometric phase.
This can be gauged away unless we consider a closed
cycle. For a closed cycle, the gauge invariant
phase $(1/\hbar) \oint \bm{A}\cdot \vec{d\bm{x}}$ is called Berry phase.

With the above zero-order solution
we can obtain the following result:
\begin{eqnarray}
\langle F^k \rangle &=&
\left\langle  n(\bm{x})   \left|-\frac{\partial {\cal H}}{\partial x_k}  \right|  n(\bm{x})  \right\rangle
\\ &=&
-\frac{\partial}{\partial x_k}
\left\langle n(\bm{x})|\ {\cal H}(\bm{x}) \ |n(\bm{x})\right\rangle
\end{eqnarray}
In case of the standard examples that were mentioned previously
this corresponds to conservative force or to persistent current.
From now on we ignore this
trivial contribution to $\langle F^k \rangle$ ,
and look for the a first order contribution.

\section{The stationary adiabatic solution: Adiabatic Transport or ``Geometric Magnetism"}

For linear driving (unlike the case of a cycle) the
$\bm{A}_n(\bm{x})$ field can be gauged away.
Assuming further that the adiabatic equation
can be treated as parameter independent
(that means disregarding the parametric dependence
of $E_n$ and ${W}$ on $\bm{x}$) one realizes
that Eq.(\ref{e_24}) possesses stationary solutions.
To first order these are:
\begin{eqnarray} \label{e_28}
| \psi \rangle \ = \
|n\rangle +
\sum_{m(\ne n)}
\frac{\mbf{W}_{mn}}
{E_n-E_m} |m\rangle
\end{eqnarray}
Note that in a {\em fixed-basis representation}
the above stationary solution is in fact time-dependent.
Hence the notations $|n(\bm{x}(t))\rangle$,  $|m(\bm{x}(t))\rangle$
and  $| \psi(t) \rangle$ are possibly more appropriate.

With the above solution
we can write $\langle F^k \rangle$ as a sum of
zero order and first order contributions.
From now on we ignore the
zero order contribution,
and go on with the first order contribution:
\begin{eqnarray}
\langle F^k \rangle &=&
-\sum_{m(\ne n)}
\frac{\mbf{W}_{mn}}{E_n-E_m}
\left\langle n \Big|
\frac{\partial {\cal H}}{\partial x_k}
\Big| m \right\rangle + \mbox{CC}
\nonumber \\
&=& \sum_j \left(i\sum_{m}
\mbf{A}^k_{nm} \mbf{A}^{j}_{mn} + \mbox{CC}\right) \ \dot{x}_j
\nonumber \\
&=& -\sum_j \mbf{B}_n^{kj} \ \dot{x}_j
\end{eqnarray}
For a general {\em stationary} preparation,
either pure or mixed, one obtains Eq.(\ref{e_8}) with
\begin{eqnarray} \label{e_30}
\fullG^{kj} \ = \ \sum_n f(E_n) \ \mbf{B}_n^{kj}
\end{eqnarray}
where  $f(E_n)$ are weighting factors,
with the normalization $\sum_n f(E_n)=1$.
For a pure state preparation $f(E_n)$ distinguishes
only one state $n$, while for canonical
preparation $f(E_n)\propto\exp(-E_n/T)$,
where $T$ is the temperature.
For a many-body system of non-interacting Fermions
$f(E_n)$ can be re-interpreted as
the Fermi occupation function,
so that $\sum_n f(E_n)$ is the total
number of particles.

Thus we see that the assumption of a stationary
first-order solution leads to a non-dissipative (antisymmetric)
conductance matrix This is know as either
Adiabatic Transport \cite{thouless,avron}
or ``Geometric Magnetism" \cite{berry}.
In the later sections we shall discuss
the limitations of the above result.

\section{The non-stationary solution: The Kubo formula}

The Kubo formula is an expression for the linear response
of a driven system that goes beyond the stationary
adiabatic solution of the previous section.
The Kubo formula has many type of derivations. 
One possibility is to use the same procedure 
as in Section~5 starting with
\begin{eqnarray} \nonumber
| \psi(t) \rangle \ &=& \
\eexp{-iE_nt}|n\rangle \\ \nonumber
&+& \sum_{m(\ne n)}
\left[-i\mbf{W}_{mn}\int_0^t\eexp{i(E_n{-}E_m)t'}dt'\right]
\eexp{-iE_mt} |m\rangle
\end{eqnarray}
For completeness we also give in Appendix~A  
a simple version of the standard derivation, which is based
on a conventional fixed-basis first-order-treatment
of the perturbation. The disadvantages are: \\
\begin{minipage}{\hsize}
\vspace*{0.1cm}
\begin{itemize}
\setlength{\itemsep}{0cm}
\item
The standard derivation does not illuminate the underlaying
physical mechanisms of the response.
\item
The stationary adiabatic limit is not manifest.
\item
The fluctuation-dissipation relation is vague.
\item
The validity conditions of the derivation are not clear:
no identification of the {\em regimes}.
\end{itemize}
\vspace*{0.0cm}
\end{minipage}
For now we go on with the conventional approach,
but in a later section we refer to the more illuminating
approach of \cite{wilk} and \cite{vrn,dsp,crs,frc}.
Then we clarify what is the regime (range of $\dot{x}$)
were the Kubo formula can be trusted \cite{crs,frc},
and what is the sub-regime where the response can be
described as non-dissipative adiabatic transport.

In order to express the Kubo formula one
introduces the following definition:
\begin{eqnarray}
K^{ij}(\tau) = \frac{i}{\hbar} \langle [F^i(\tau),F^j(0)]\rangle
\end{eqnarray}
We use the common interaction picture notation
\mbox{$F^k(\tau)= \eexp{i{\cal H}t} F^k \eexp{-i{\cal H}t}$},
where ${\cal H}={\cal H}(\bm{x})$ with $x=\mbox{const}$.
The expectation value assumes that
the system is prepared in a {\em stationary} state
(see previous section).
It is also implicitly assumed that the result is
not sensitive to the exact value of $x$.
Note that $K^{ij}(\tau)$ has
a well defined classical limit.
Its Fourier transform will be denoted
$\tilde{K}^{ij}(\omega)$.

The expectation value $\langle F^{k} \rangle$
is related to the driving $\bm{x}(t)$ by
the causal response kernel \mbox{$\alpha^{ij}(t-t')$}.
The Kubo expression for this response kernel,
as derived in Appendix~A, is
\begin{eqnarray} \label{e_kubo}
\alpha^{ij}(\tau) \ \ = \ \
\Theta(\tau) \ K^{ij}(\tau)
\end{eqnarray}
where the step function  $\Theta(\tau)$ cares
for the upper cutoff of the integration in Eq.(\ref{e_3}).
The Fourier transform of  $\alpha^{ij}(\tau)$
is the generalized susceptibility $\chi^{ij}(\omega)$.
The conductance matrix is defined as:
\begin{eqnarray} \label{e_33}
\fullG^{ij} \ = \ \lim_{\omega\rightarrow 0}
\frac{\im[\chi^{ij}(\omega)]}{\omega} \ = \
\int_0^{\infty} K^{ij}(\tau)\tau d\tau
\end{eqnarray}
This can be split into symmetric
and anti-symmetric components
(see derivation in Appendix~C)
as follows:
\begin{eqnarray} \label{e_34}
\symG^{ij} = \half(\fullG^{ij}{+}\fullG^{ji}) &=&
\frac{1}{2}\lim_{\omega\rightarrow 0}
\frac{\im[\tilde{K}^{ij}(\omega)]}{\omega}
\\ \label{e_35}
\mbf{B}^{ij} = \half(\fullG^{ij}{-}\fullG^{ji}) &=&
{-}\int_{-\infty}^{\infty}\frac{d\omega}{2\pi}
\frac{\re[\tilde{K}^{ij}(\omega)]}{\omega^2}
\end{eqnarray}
The antisymmetric part is identified \cite{robbins} (Appendix~C)
as corresponding to the stationary solution Eq.(\ref{e_30})
of the adiabatic equation.

\section{The emergence of dissipative response, and the fluctuation-dissipation relation}

The Kubo formula for the symmetric part of the
conductance matrix ($\symG^{ij}$) can be further simplified.
If we take Eq.(\ref{e_34}) literally, then $\symG^{ij}=0$ due to
the simple fact that we have finite spacing between
energy levels (see \cite{ophir} for a statistical point of view).
But if we assume that the energy levels
have some finite width $\Gamma$, then the smoothed
version of $\tilde{K}^{ij}(\omega)$ should
be considered. In common textbooks the introduction
of $\Gamma$ is ``justified" by assuming
some weak coupling to an environment,
or by taking the limit of infinite volume.
But we are dealing with a strictly isolated finite
system, and therefore the meaning of $\Gamma$
requires serious consideration.
We postpone the discussion of this issue
to the next section.

If the smoothed version of $\tilde{K}^{ij}(\omega)$
should be used in Eq.(\ref{e_34}), then it is possible to obtain
$\symG^{ij}$ from power spectrum  $\tilde{C}^{ij}(\omega)$
of the fluctuations. This is called the Fluctuation-Dissipation relation.
The spectral function $\tilde{C}^{ij}(\omega)$
is defined as the Fourier transform of the
symmetrized correlation function
\begin{eqnarray}
C^{ij}(\tau) = \langle \half (F^i(\tau)F^j(0)+F^j(0)F^i(\tau)) \rangle
\end{eqnarray}
We use again the interaction picture,
as in the definition of $K^{ij}(\tau)$.
Also this function has a well defined classical limit.

There are several versions for the
Fluctuation-Dissipation relation.
The microcanonical version \cite{robbins}
has been derived using classical considerations,
leading to
\begin{eqnarray} \label{e43}
\symG^{ij}|_E \ = \
\frac{1}{2}\frac{1}{g(E)}
\frac{d}{dE}\left[g(E) \ \tilde{C}_E^{ij}(\omega\rightarrow0)\right]
\end{eqnarray}
In Appendix~C we introduce its quantum mechanical derivation.
The subscript emphasizes that we assume
a microcanonical state with energy $E$,
and $g(E)$ is the density of states.
The traditional version of the Fluctuation-Dissipation relation
assumes a canonical state. It can be obtained
by canonical averaging over the microcanonical version leading to:
\begin{eqnarray}
\symG^{ij}|_T \ = \
\frac{1}{2T}\tilde{C}_T^{ij}(\omega\rightarrow0)
\end{eqnarray}

\section{The validity of linear response theory and beyond}

The standard derivation of the dissipative part of the Kubo formula,
leading to the Fluctuation-Dissipation relation Eq.(\ref{e43}),
is not very illuminating physically.
More troubling is the realization that one cannot tell
from the standard  derivation what are the conditions for its validity.
An alternate, physically appealing derivation \cite{vrn,dsp,crs,frc},
is based on the observation that energy absorption is related
to having diffusion in energy space \cite{wilk}.
In Appendix~E we outline the main ingredients of this approach.

It should be clear that the diffusion picture of Appendix~E
holds only in case of {\em chaotic systems}. If this diffusion picture
does not hold, then also the Kubo formula for $\symG^{ij}$ does not hold!
Driven one-dimensional systems are the obvious example
for the failure of linear response theory (LRT).
As in the case of the kicked rotator (standard map) \cite{qkr}
there is a complicated route to chaos and stochasticity:
By increasing  the driving amplitude the phase space
structure is changed. If the amplitude is smaller than
a threshold value, then the diffusion is blocked by Kolmogorov-Arnold-Moser curves,
and consequently there is not dissipation.
Therefore the Kubo formula is not applicable in such cases.

The following discussion of {\em dissipative response}
assumes that we deal with a quantized chaotic system.
We would like to discuss the reason and the consequences
of having an energy scale $\Gamma$. In the standard derivation
the assumption of having level broadening as if comes out of the blue.
As we already noted it is customary in textbooks to argue
that either a continuum limit, or some small coupling to an environment
is essential in order to provide $\Gamma$. But this is of course
just a way to avoid confrontation with the physical  problem
of having a driven {\em  isolated finite mesoscopic system}.
In fact the energy scale $\Gamma$ is related to the rate ($\dot{x}$)
of the driving:
\begin{eqnarray} \label{e_42}
\Gamma \ \ = \ \
\left(\frac{\hbar\sigma}{\Delta^2}|\dot{x}|\right)^{2/3} \times \Delta
\end{eqnarray}
where for simplicity of presentation we assume
one parameter driving. We use $\Delta$ to denote the
mean level spacing, and $\sigma$ is the root mean
square value of the matrix element $F_{nm}$
between neighboring levels.
In order to derive the above expression for $\Gamma$
we have used the result of \cite{frc} (Sec.17) for the ``core width"
at the breaktime $t=t_{\tbox{prt}}$ of perturbation theory.
The purpose of the present section is to give an optional
"pedestrian derivation" for $\Gamma$, and to discuss
the physical consequences.

Looking at the first order solution Eq.(\ref{e_28}) of Section~5
one realizes that it makes sense provided $|\mbf{W}_{mn}| \ll \Delta$.
This leads to the adiabaticity condition
\begin{eqnarray} \label{e91}
|\dot{x}| \ \ \ll \ \ \frac{\Delta^2}{\hbar\sigma}
\end{eqnarray}
If this condition is not satisfied one should
go beyond first order perturbation theory (FOPT),
in a sense to be explained below.
Note that this adiabaticity condition can be
written as $\Gamma\ll\Delta$.

The adiabaticity condition Eq.(\ref{e91}) can be explained
in a more illuminating way as follows:
Let us assume that we prepare the system at time $t=0$
at the level $|n\rangle$.
Using {\em time dependent} FOPT we find out that
a stationary-like solution is reached after
the Heisenberg time $t_{\tbox{H}}=2\pi\hbar/\Delta$.
This is of course a valid description
provided we do not have by then a breakdown
of FOPT. The condition for this is
easily found to be $\dot{x}t_{\tbox{H}} \ll \delta x_c$,
where $\delta x_c=\Delta/\sigma$. This
leads again to the adiabaticity condition Eq.(\ref{e91}).

Another assumption in the derivation of
Section~5 was that we can
ignore the parametric dependence
of $E_n$ and ${W}$ on $\bm{x}$.
The adiabaticity condition
$t_{\tbox{H}} \ll \delta x_c/\dot{x}$
manifestly justify such an assumption:
We should think of $t_{\tbox{H}}$ as
the transient time for getting
a stationary-like state, and we
should regard $\delta x_c$ as the parametric
correlation scale.

As strange as it sounds, in order to
have dissipation, it is essential to have
a breakdown of FOPT. In the language
of perturbation theory this implies
a required summation of diagrams to {\em infinite order},
leading to an effective broadening of the energy levels.
By iterating FOPT, neglecting interference terms,
we get a {\em Markovian approximation}
for the energy spreading process.
This leads to the diffusion equation of Appendix~E.
This diffusion can be regarded as arising
from Fermi-golden-rule transitions between
energy levels.
A simple ad-hoc way to determine the
energy level broadening  is to introduce  $\Gamma$
as a lower cutoff in the energy distribution
which is implied by Eq.(\ref{e_28}):
\begin{eqnarray}
|\langle n | \psi \rangle|^2 \ = \
\frac{|\hbar \dot{x} \mbf{F}_{mn}|^2}
{(E_n-E_m)^4 + (\Gamma)^4}
\end{eqnarray}
This constitutes a generalization of the well known
procedure used by Wigner in order to obtain
the local density of states \cite{vrn,dsp}.
However, in the present context we do not get a Lorentzian.
The width parameter $\Gamma$ is determined
self consistently from normalization,
leading to Eq.(\ref{e_42})  [disregarding numerical prefactor].

We can summarize the above reasoning by saying
that there is a {\em perturbative regime}
that includes an adiabatic (FOPT) sub-regime.
Outside of the adiabatic sub-regime we need {\em all orders}
of perturbation theory leading to Fermi-golden-rule
transitions, diffusion in energy space,  and hence dissipation.
Thus the dissipative part of Kubo formula
emerges only in the regime $\Gamma>\Delta$,
which is just the opposite of the adiabaticity condition.
The next obvious step is to determine
the boundary of the perturbative regime.
Following \cite{crs,frc} we argue
that the required condition is $\Gamma\ll\Delta_b$.
The bandwidth $\Delta_b\propto\hbar$
is defined as the energy width
$|E_n-E_m|$ were the matrix
elements $F_{nm}$ are not vanishingly small.
If the condition $\Gamma\ll\Delta_b$
is violated we find ourselves
in the non-perturbative regime 
where the Kubo formula cannot be trusted \cite{crs,frc}.

We still have to illuminate why we can
get in the perturbative regime
a dissipative {\em linear} response 
in spite of the breakdown of FOPT.
The reason is having a separation of scales
($\Delta \ll \Gamma\ll\Delta_b$).
The non-perturbative mixing on
the small energy scale $\Gamma$ does not
affect the rate of first-order
transitions between distant levels
\mbox{($\Gamma \ll |E_n-E_m| \ll \Delta_b$)}.
Therefore Fermi golden rule picture
applies to the description
of the coarse grained energy spreading,
and we get linear response.

The existence of the adiabatic regime is obviously
a quantum mechanical effect. If we take
the formal limit $\hbar\rightarrow 0$
the adiabaticity condition  $\Gamma \ll \Delta$  breaks down.
In fact the {\em proper} classical limit
is {\em non-perturbative},
because also the weaker condition
$\Gamma \ll \Delta_b$ does not survive
the $\hbar\rightarrow 0$ limit.
For further details see \cite{vrn,dsp,crs,frc}.
In the non-perturbative regime the
quantum mechanical derivation of Kubo formula
is not valid. Indeed we have demonstrated \cite{crs}
the failure of Kubo formula in case of
random-matrix models. But if the system
has a classical limit, then Kubo formula
still holds in the non-perturbative regime
due to {\em semiclassical} (rather than quantum-mechanical)  reasons.

The discussion of {\em dissipation} assumes
a generic situation such that
the Schrodinger equation does not have
a stationary solution. This means that
driven one-dimensional systems are automatically
excluded. Another non-generic possibility
is to consider a special driving scheme,
such as translation, rotation or dilation \cite{dil}.
In such case the time dependent Hamiltonian
${\cal H}(\bm{x}(t))$  possesses a stationary solution
(provided the ``velocity" $\dot{x}$
is kept constant). Consequently we do
not have a dissipation effect.
In Section~11 we discuss the simplest
example of pumping by translation, where
the stationary adiabatic solution of Section~5
is in fact exact, and no dissipation arises.

\section{Application to pumping}

So far we have discussed the response for driving in a very
general way. From now on we focus on a system with a ring
geometry as described in the Introduction, and illustrated
in Fig.~1.  The shape of the ring is controlled by some
parameters $x_1$ and $x_2$, and $x_3$ is the magnetic flux.
The generalized force $F^3$ which is conjugate
to the flux is the current. The time integral over the
current is the transported charge:
\begin{eqnarray}
Q \ \ = \ \ \oint \langle F^{3} \rangle dt
\end{eqnarray}
In fact a less misleading terminology is to talk
about ``probability current" and
"integrated probability current".
From a purely mathematical point of view
it is not important whether the transported
particle has an electrical charge.

Disregarding a possible  persistent current contribution,
the expression for the pumped charge is:
\begin{eqnarray} \label{e6}
Q \ = \  -\left[ \oint \bmsf{G} \cdot d\bm{x} + \oint \bmsf{B} \wedge d\bm{x} \right]_{k=3}
\end{eqnarray}
If we neglect the first term, which is associated
with the dissipation effect, and average the second
("adiabatic") term over the flux, then we get
\begin{eqnarray} \label{e7}
\overline{Q|_{\tbox{adiabatic}}} \ = \
-\frac{1}{2\pi\hbar}\intt \bmsf{B} \cdot \vec{d\bm{x}} \wedge \vec{d\bm{x}}
\end{eqnarray}
The integration should be taken over
a cylinder of vertical height $2\pi\hbar$,
and whose basis is determined by the projection of
the pumping cycle onto the $(x_1,x_2)$ plane.

We already pointed out that the Berry phase
$(1/\hbar)\oint  \bm{A}_n\cdot \vec{d\bm{x}}$
is gauge invariant. Therefore from Stokes law it follows that
$(1/\hbar)\intt \bmsf{B}  \cdot \vec{d\bm{x}} \wedge \vec{d\bm{x}}$
is independent of the surface, and therefore
$(1/\hbar){\ointt} \bmsf{B}  \cdot \vec{d\bm{x}} \wedge \vec{d\bm{x}} $
with closed surface should be $2\pi\times$integer.
Integrating over a cylinder, as in Eq.(\ref{e7}),
is effectively like integrating over a closed surface
(because of the $2\pi$~periodicity in the vertical
direction). This means that the flux averaged $Q$
of Eq.(\ref{e7}) has to be an integer.

The common interest is in pumping cycles
in the $\Phi=0$ plane. This means  that the
zero order conservative  contribution to~$Q$,
due to  a persistent current, does not exist.
Furthermore, from the reciprocity relations
(see Appendix~B)  it follows that
$\fullG^{31}=- \fullG^{13}$,
and $\fullG^{32}=- \fullG^{23}$,
which should be contrasted with
 $\fullG^{12}=\fullG^{21}$.
This means that a pumping cycle in the
$\Phi=0$ plane is purely adiabatic:
there is no dissipative contribution to~$Q$.
Only the $\vec{\bm{B}}$ field   (second term in Eq.(\ref{e6}))
is relevant to the calculation of the pumped charge,
and its vertical component ${B}^{12}$
vanishes due to the time reversal symmetry.

The absence of dissipative contribution for
a cycle in the $\Phi=0$ plane, does not
imply that dissipation is not an issue.
The symmetric part of the conductance
matrix $\symG^{ij}$ is in general non-zero,
leading to an energy absorption
rate which is proportional to $\dot{x}^2$.
This implies that the energy absorption per cycle
is proportional to $|\dot{x}|$. Therefore we
are able to minimize the dissipation effect by making
the pumping cycle very slow. Furthermore,
if we get into the quantum-mechanical adiabatic regime,
then $\symG^{ij}$ becomes extremely small, and then
we can neglect the dissipation effect as long as
quantum-mechanical adiabaticity can be trusted.

Whenever the dissipation effect cannot be neglected,
one should specify whether or how a {\em stationary operation} is achieved.
In case of pumping in open system the stationary
operation is implicitly guaranteed by having equilibrated reservoirs,
where the extra energy is dissipated to infinity.
In case of pumping in closed system the issue of stationary
operation is more subtle:
In the adiabatic regime, to the extend that adiabaticity
can be trusted, we have a stationary solution to the transport problem,
as defined in Section~5.
But outside of the adiabatic regime
we have diffusion in energy space (Appendix~E)
leading to a slow energy absorption (dissipation).
Thus a driven system  is heated up gradually
(though possibly very slowly).  Strictly speaking
a stationary operation is not achieved,
unless the system is in (weak)  thermal contact
with some large bath.
Another way to reach a stationary operation,
that does not involve an external bath, 
is by having an effectively bounded phase space.
This is the case with the mixed phase space example
which is discussed in Ref.\cite{ratchets}.
There the stochastic-like motion takes place
in a bounded chaotic region in phase space.

\section{classical dissipative pumping}

Before we discuss the quantum mechanical pumping,
it is instructive to bring simple examples for
{\em classical} pumping. In the following we consider
one particle  ($\bm{r}$)
in a two dimensional ring as in Fig.1a.

The first example is for classical {\em dissipative} pumping.
The conductance $G=\fullG^{33}$
can be calculated for this system \cite{wlf}
leading to a mesoscopic variation of the Drude formula.
The current is given by Ohm law $I=-G\times\dot{\Phi}$, where $-\dot{\Phi}$
is the electro-motive-force.

Consider now the following pumping cycle:
Change the flux from $\Phi_1$ to $\Phi_2$,
hence pumping charge \mbox{$Q = -G(1)\times (\Phi_2-\Phi_1)$}.
Change the conductance from $G(1)$ to $G(2)$
by modifying the shape of the ring.
Change the flux from $\Phi_2$ back to $\Phi_1$,
hence pumping charge \mbox{$Q(2) = -G(2) \times (\Phi_1-\Phi_2)$}.
Consequently the net pumping is
\begin{eqnarray}
Q \ = \ (G(2)-G(1)) \times (\Phi_2-\Phi_1)
\end{eqnarray}
Thus we have used the dissipative part of the conductance
matrix (first term in Eq.(\ref{e6})) in order to pump charge.
In the quantum mechanical  version of this example
extra care should be taken with respect to the
zero order contribution of the persistent current.

\section{classical adiabatic pumping}

The second example is for classical {\em adiabatic} pumping.
The idea is to trap the particle
inside the ring by a potential well,
and then to make a translation
of the trap along a circle. The result of such a cycle
is evidently $Q=1$. We would like to see
how this trivial result emerges form the Kubo formula.

Let $(\bm{r},\bm{p})$ be the canonical coordinate of the particle in the ring,
while $(x_1,x_2)$ are the center coordinate of a trapping potential.
The Hamiltonian is:
\begin{eqnarray}
{\cal H}(\bm{r},\bm{p};\bm{x}(t))  &=&
\frac{1}{2\mathsf{m}}
\left[
\mbf{p}_{\perp}^2 +
\left(\mbf{p}_{\parallel}-\frac{1}{2\pi\sqrt{x_1^2+x_2^2}}\Phi(t)\right)
\right] \nonumber \\ &+&
U_{\tbox{trap}}(\mbf{r}_1{-}x_1(t),\mbf{r}_2{-}x_2(t))
\end{eqnarray}
where $\mbf{p}_{\parallel}$ and $\mbf{p}_{\perp}$ are the components
of the momentum along the ring and in the perpendicular (transverse) directions.
The pumping is done simply by cycling the position of the trap.
The translation of the trap is assumed to be along an inside circle
of radius $R$,
\begin{eqnarray}
\bm{x}(t) = (R\cos(\Omega t),R\sin(\Omega t),\Phi{=}\const)
\end{eqnarray}

In this problem the stationary solution of Section~5
is an exact solution. Namely
\begin{eqnarray} \label{e_psit}
|\psi(t)\rangle \ = \eexp{i\mathsf{m}\dot{\bm{x}}\cdot\bm{r}} \ |n(\bm{x}(t))\rangle
\end{eqnarray}
where $|n(\bm{x})\rangle\mapsto \psi^{(n)}(\bm{r}-\bm{x})$ are
the eigenfunctions of a particle in the trap.
Eq.(\ref{e_psit}) is just Galilei transformation from
the moving (trap) frame to the Laboratory frame.

It is a-priori clear that in this
problem the pumped charge per cycle is $Q=1$,
irrespective of $\Phi$. Therefore the $\vec{\bm{B}}$ field
must be
\begin{eqnarray} \label{e_52}
\vec{\bm{B}} = -\frac{(x_1,x_2,0)}{2\pi (x_1^2+x_2^2)}
\end{eqnarray}
This can be verified by calculation via Eq.(\ref{e29}).
The singularity along the $x_3$ axis
is not of quantum mechanical origin:
It is not due to degeneracies, but rather
due to the diverging current operator
($\partial {\cal H}/\partial x_3\propto 1/\sqrt{x_1^2+x_2^2}$).

\section{quantum pumping}

We turn now to the quantum mechanical case.
Consider an adiabatic cycle that
involves a particular energy level $n$.
This level is assumed to have a degeneracy
point at $(x_1^{(0)},x_2^{(0)},\Phi^{(0)})$.
It follows that in fact there is
a vertical chain of degeneracy points:
\begin{eqnarray} \label{e_chain}
\mbox{chain} \ = \ (x_1^{(0)},x_2^{(0)},\Phi^{(0)}+2\pi\hbar\times\mbox{\small integer})
\end{eqnarray}
These degeneracy points are important for the geometrical
understanding of the $\vec{\bm{B}}$ field, as implied by Eq.(\ref{e29}).
Every degeneracy point is like a monopole charge.
The total flux that emerges from each monopole
must be  $2\pi\hbar\times$integer for a reason that was
explained after Eq.(\ref{e7}).
Thus the monopoles are quantized in units of~$\hbar/2$.
The $\vec{\bm{B}}$ field which is created (so to say)
by a vertical chain of monopoles may have a different {\em near field}
and {\em far field} behavior, which we discuss below.

The far field region exists if the chains are
well isolated.  Later we explain that ``far"
means $g_T \ll 1$, where $g_T$ is the Thouless conductance.
The far field  is obtained by regarding the chain as a smooth line.
This leads {\em qualitatively} to the same field as in Eq.(\ref{e_52}).
Consequently, for a ``large radius" pumping cycle
in the $\Phi=0$ plane, we get $|Q|\approx1$.
In the following we are interested in
the deviation from exact quantization:
If $\phi^{(0)}=0$ we expect to have
$|Q| \ge 1$, while if $\phi^{(0)}=\pi$ we expect $|Q|\le 1$.
Only for the $\phi$ averaged $Q$ of Eq.(\ref{e7})
we get {\em exact quantization}.

The deviation from $|Q|\approx 1$ is extremely large
if we consider a tight pumping cycle around
a $\phi^{(0)}=0$ degeneracy.
After linear transformation of the shape parameters,
the energy splitting $\Delta=E_n-E_m$ of the energy level~$n$
from its neighboring (nearly degenerated) level~$m$
can be written as
\begin{eqnarray}
\Delta=
((x_1{-}x_1^{(0)})^2+(x_2{-}x_2^{(0)})^2+
c^2(\phi{-}\phi^{(0)})^2)^{1/2}
\end{eqnarray}
where $c$ is a constant. The monopole field is accordingly
\begin{eqnarray} \label{e9}
\vec{\bm{B}} = \pm\frac{c}{2} \
\frac{( x_1{-}x_1^{(0)},  x_2{-}x_2^{(0)},   x_3{-}x_3^{(0)}) }
{((x_1{-}x_1^{(0)})^2+(x_2{-}x_2^{(0)})^2+
(\frac{c}{\hbar})^2(x_3{-}x_3^{(0)})^2)^{3/2}}
\end{eqnarray}
where the prefactor is determined by the requirement
of having a single ($\hbar/2$) monopole charge.
Assuming a pumping cycle of radius $R$ in the $\Phi=0$ plane
we get from the second term of Eq.(\ref{e6})
\begin{eqnarray}
Q \ = \  -\left[\oint \bmsf{B} \wedge d\bm{x} \right]_3
\ = \  \mp \pi \sqrt{g_T}
\end{eqnarray}
where
\begin{eqnarray}
g_T \ = \
\frac{1}{\Delta}
\frac{\partial^2\Delta}{\partial\phi^2} \ = \
\frac{c^2}{R^2}
\end{eqnarray}
is a practical definition for the Thouless conductance
in this context. It is used here simply as a measure
for the sensitivity of an energy level to the magnetic flux $\Phi$.

What we want to do in the next sections is to interpolate
between the near field result, which is  $Q={\cal O}(\sqrt{g_T})$,
and the far field  result, which is $Q={\cal O}(1)$.
For this purpose it is convenient to consider
a particular model that can be solved exactly.

\section{The double barrier model}

A simple example for quantum pumping is the double barrier model.
An open geometry version of this model has been
analyzed in \cite{barriers} using the $S$~matrix approach.
The analogous closed geometry version is obtained
by considering a one-dimensional  ring with two delta barriers.
As we are going to explain below,
the pumping process in this model can be regarded
as a particular example of an adiabatic transfer scheme:
The electrons are adiabatically transfered from
state to state one by one as in ``musical chair game".

The two delta barriers version of the double barrier model
is illustrated in Fig.2. The length of the ring is $L$,
with periodic boundary conditions on $-(L/2)<r<(L/2)$.
A dot region $|Q|<a/2$ is defined by the potential
\begin{eqnarray}
U(r;c_1,c_2) =
\frac{1}{c_1}\delta\left(r+\frac{a}{2}\right) +
\frac{1}{c_2}\delta\left(r-\frac{a}{2}\right)
\end{eqnarray}
It is assumed that $c_1$ and $c_2$ are small enough
so one can classify the ring eigenstates into two
categories: wire states, and dot states. The
latter are those states that are localized in the
dot region $|Q|<a/2$ in the limit of infinitely
high barriers. We define the Fermi energy as
the energy of the last occupied wire
level in the limit of infinitely high barriers.

The three parameters that we can control
are the flux $x_3 = \Phi = \hbar\phi$,
the bias $x_1=c_1-c_2$,
and the dot potential $x_2=E_{\tbox{dot}}$
which is related to  $c_1+c_2$.
The energy $E_{\tbox{dot}}$ correspond
to the dot state which is closest to the Fermi
energy $E_F$ from above.
We assume that the other dot levels
are much further away from the Fermi energy,
and can be ignored. Note that another possible
way to control the dot potential, is simply
by changing a gate voltage: That means to
assume that there is a control over the
potential floor in the region $|Q|<a/2$.

{\em The pumping cycle is assumed to be in the
$\Phi=0$ plane, so there is no issue
of conservative persistent current contribution}.
The pumping cycle is defined as follows:
We start with a positive bias ($x_1>0$)
and lower the dot potential from a large $x_2>E_F$
value to a small $x_2<E_F$ value.
As a result, one electron is
transfered via the {\em left} barrier into the
dot region. Then we invert the bias
($x_1<0$) and raise back $x_2$. As a result
the electron is transfered back into
the wire via the {\em right} barrier.

A closer look at the above scenario (Fig.2b)
reveals the following:
As we lower the dot potential across a wire
level, an electron is adiabatically transfered
once from left to right and then from right
to left. As long as the bias is
positive ($x_1>0$) the net charge being pumped
is very small ($|Q| \ll 1$). Only the lowest wire
level that participate in the pumping cycle
carries $Q={\cal O}(1)$ net charge:
It takes an electron from the left side,
and after the bias reversal it emits it
into the right side.
Thus the pumping process in this model can be regarded
as a particular example \cite{avron} of an {\em adiabatic transfer scheme}:
The electrons are adiabatically transfered from
state to state, one by one, as in ``musical chair game".

For a single occupied level the net $Q$ is the
sum of charge transfer events that take place in
four avoided crossings (two avoided crossings
in case of the lowest level).
For many particle occupation
the total $Q$ is the sum over the net $Q$s
which are carried by individual levels.
For a dense zero temperature Fermi occupation
the summation over all the net $Q$s is a telescopic sum,
leaving non-canceling contributions only from the
first and the last adiabatic crossings. The latter
involve the last occupied level at the Fermi energy.

\section{The three site lattice Hamiltonian}

Rather than analyzing the two-delta-barriers version
of the double barrier model, we consider below
a simplified version that still contains the {\em same} essential
ingredients. This is obtained by considering a three site lattice Hamiltonian.
The advantage is obviously the possibility to make an exact analytical
treatment that does not involve approximations.

The middle site in the three site lattice Hamiltonian
supports a single dot state,
while the two other sites support
two wire states. The Hamiltonian is
\begin{eqnarray}
{\cal H} \mapsto \left(
\matrix{
0 & c_1 & \eexp{i\phi} \cr
c_1 & u & c_2 \cr
\eexp{-i\phi} & c_2 & 0} \right)
\end{eqnarray}
The three parameters are
the bias $x_1=c_1-c_2$,
the dot energy $x_2=u$,
and the flux $x_3 = \Phi = \hbar\phi$.
For presentation purpose we assume
that \mbox{$0 < c_1,c_2 \ll 1$},
and characterize the wire-dot coupling
by the parameter $c=\sqrt{c_1c_2}$.

The eigenstates are $E_n$.
Disregarding the interaction
with the dot ($c=0$) we have two
wire states with  \mbox{$E=\pm1$}.
This implies degeneracies
for $x_2=u=\mp1$.
Once we switch on the coupling ($c>0$),
the only possible degeneracies are between
the even dot state and the odd wire state
of the mirror symmetric Hamiltonian ($x_1=0$).
The flux should be either integer
(for degeneracy of the dot level with the lower wire level),
or half integer
(for degeneracy of the dot level with the upper wire level).
Thus we have two vertical chains of degeneracies:
\begin{eqnarray} \nonumber
\mbox{The negative chain}
\ &=& \  (0,-1{+}c^2,2\pi\hbar\times\mbox{\small integer})
\\ \nonumber
\mbox{The positive chain}
\ &=& \ (0,+1{-}c^2,\pi+2\pi\hbar\times\mbox{\small integer})
\end{eqnarray}

In order to calculate the $\vec{\bm{B}}$ field and pumped
charge $Q$, we have to find the eigenvalues
and the eigenvectors of the Hamiltonian matrix.
The secular equation for the eigenvalues is
\begin{eqnarray} \nonumber
E^3 - u E^2 - (1+c_1^2+c_2^2)E
+ u - 2 c_1 c_2 \cos(\phi) = 0
\end{eqnarray}
Using the notations
\begin{eqnarray} \nonumber
{\cal Q} &=&  \frac{1}{9}u^2 +  \frac{1}{3}(1+c_1^2+c_2^2)
\\ \nonumber
{\cal R} &=&  \frac{1}{27}u^3
+ \frac{1}{6}(1+c_1^2+c_2^2)u - \frac{1}{2}u
+ c_1c_2 \cos(\phi)
\\ \nonumber
& & \cos(\theta)=\frac{{\cal R}}{\sqrt{{\cal Q}^3}}
\end{eqnarray}
the roots of the above cubic equation are:
\begin{eqnarray}
E_n = \frac{1}{3}u + 2\sqrt{{\cal Q}}\cos\left(\frac{1}{3} \theta  + n\frac{2\pi}{3} \right)
\end{eqnarray}
where $n=0,\pm1$.
The corresponding eigenstates are:
\begin{eqnarray}
|n(\bm{x})\rangle \mapsto
\frac{1}{\sqrt{S}}
\left(
\matrix{
c_2 \eexp{i\phi} + c_1 E_n \cr
1 - E_n^2 \cr
c_1\eexp{-i\phi} + c_2 E_n} \right)
\end{eqnarray}
where $S$ is the normalization, namely
\begin{eqnarray}
S =  (1 {-} E_n^2)^2
+ (c_1 {+} c_2 E_n)^2
+ (c_2 {+} c_1 E_n)^2
\end{eqnarray}
For the calculation of the
pumped charge in the next paragraph
it is useful to notice that for $E=\pm 1$
the normalization is $S=2(c_1 \pm c_2)^2$,
while for $E=0$ the normalization is $S\approx 1$.

After some algebra we find that
the first component of the
$\vec{\bm{B}}$ field in the
$\Phi=0$ plane is
\begin{eqnarray}
{\mbf{B}}^1 &=&
-2\Im\left.\left\langle\frac{\partial}{\partial u} n(\bm{x}) \right|
\frac{\partial}{\partial \phi} n(\bm{x}) \right\rangle
\\ \ &=&
-(c_1^2-c_2^2)
\frac{1}{S^2}
\frac{\partial S}{\partial u}
\end{eqnarray}
Which is illustrated in Fig.3.
From here it follow that if we
keep constant bias, and change
only $x_2=u$, then the pumped charge is:
\begin{eqnarray}
Q = -\int {\mbf{B}}^1 dx_2 =
\left.
- (c_1^2-c_2^2)\frac{1}{S}
\right|_{\tbox{initial}}^{\tbox{final}}
\end{eqnarray}
For a planar ($\Phi=0$) pumping cycle around
the negative vertical chain the main contribution
to $Q$ comes from the two crossings
of the $x_2 \approx -1$ line.
Hence we get
\begin{eqnarray} \label{e_Qmodel}
Q = \frac{c_1+c_2}{c_1-c_2} =
\sqrt{1+2g_T}
\end{eqnarray}
where the Thouless conductance
in this context refers to the
avoided crossing, and is defined as
\begin{eqnarray}
g_T \equiv \left.\frac{1}{\Delta}
\frac{\partial^2\Delta}
{\partial\phi^2}\right|_{\phi=0}
= \frac{2c_1c_2}{(c_1-c_2)^2}
\end{eqnarray}
A similar calculation of the pumped charge for
a planar cycle around the positive chain leads to
\begin{eqnarray}
Q = -\frac{c_1-c_2}{c_1+c_2} =
-\sqrt{1-2g_T}
\end{eqnarray}
with $g_T = {2c_1c_2}/{(c_1 + c_2)^2}$.
In both cases we have approximate quantization
$Q=\pm 1 + {\cal O}(g_T)$ for $g_T \ll 1$,
while for a tight cycle either $Q \rightarrow \infty$
or $Q \rightarrow 0$ depending on which
line of degeneracies is being encircled.
If the pumping cycle encircles both chains then
we get $Q = {4c_1c_2}/{(c_1^2-c_2^2)}$.
In the latter case $Q={\cal O}(g_T)$ for $g_T \ll 1$,
with no indication for quantization.

\section{Summary and Discussion}

We have shown how the Kubo formalism can
be used in order to derive both classical and quantum
mechanical results for the pumped charge $Q$ in
a closed system. In this formulation the distinction
between dissipative and non-dissipative contributions is manifest.

Within the framework of the Kubo formalism (disregarding non-linear corrections)
we have made a distinction between the following levels of treatment: \\
\begin{minipage}{\hsize}
\vspace*{0.1cm}
\begin{itemize}
\setlength{\itemsep}{0cm}
\item
Strict adiabaticity \\ (outcome of zero order treatment)
\item
Adiabatic transport \\ (outcome of stationary first order treatment)
\item
Dissipation \\ (the result of first order transitions)
\end{itemize}
\vspace*{0.0cm}
\end{minipage}
In the adiabatic regime one can assume
a {\em stationary} solution to the adiabatic equation,
which implies no dissipation effect.
This leads to the picture of adiabatic transport,
where the Berry phase is the outcome of a zero order treatment,
while the ``geometric magnetism" of Eq.(\ref{e_30})  is the outcome
of a first order treatment of the inter-level couplings.

In some very special cases
(translations, rotations and dilations)
this assumption (of having a stationary solution)
is in fact exact, but in generic circumstance this
assumption is an approximation.
Outside of the adiabatic regime the stationary solution
cannot be trusted.

Assuming quantized {\em chaotic dynamics} one
argues that Fermi-golden-rule transitions
between levels lead to (slow) diffusion
in energy (Eq.(\ref{eD1})). This leads to the emergence
of the dissipative part in the Kubo formula.
We have obtained an expression (Eq.(\ref{e_42}))
for the energy scale $\Gamma \propto |\dot{x}|^{2/3}$
that controls the dissipative effect.
We have explained that the dissipative contribution
to the Kubo formula is valid only in the regime
$\Delta < \Gamma \ll \Delta_b$. Otherwise the dynamics
is either of adiabatic nature ($\Gamma \ll \Delta$) or
non-perturbative ($\Gamma>\Delta$).


In order to calculate the pumped charge~$Q$ we have
to perform a closed line integral over the conductance (Eq.(\ref{e6})).
This may have in general both adiabatic and dissipative
contributions. For the common pumping cycle in the
$\Phi=0$ plane, only the adiabatic contribution exists.
This follows from the reciprocity relations (Section~9).
Still we have emphasized (without any contradiction)
that in the same circumstances a dissipation effect
typically accompanies the pumping process.

The quantum adiabatic contribution
to the pumping is determined
by a line integral over a $\vec{\bm{B}}$ field
which is created by {\em monopoles}.
The monopoles, which are related
to the degeneracies of the Hamiltonian,
are located along vertical chains in $x$~space (Eq.(\ref{e_chain})).
The 3~site model provides the simplest
example for such vertical chains:
By calculating the $\vec{\bm{B}}$ field which
is created (so to say) by these chains,
we were able to determine the charge
which is pumped during a cycle (e.g. Eq.(\ref{e_Qmodel})).

The (monopoles of the) vertical chains
have {\em near field} regions (Eq.(\ref{e9})).
If the chains  are well isolated in $x$ space,
then there are also {\em far field} regions.
The far field regions are defined as those where
the Thouless conductance is very small ($g_T \ll 1$).
Pumping cycles that are contained in the
far field region of a given chain lead to
an approximately quantized pumping
\mbox{$Q = \mbox{\small integer}+{\cal O}(g_T)$}.
It is important to realize that the existence
of far field regions in $x$~space is associated with having
a low dimensional system far away from the classical limit.
In a quantized chaotic system it is unlikely
to have $g_T \ll 1$ along a pumping cycle.
As we take the $\hbar\rightarrow0$ limit the
vertical chains become very dense,
and the far field regions disappear.

In the subtle limiting case of open geometry
we expect to get agreement with the
$S$-matrix formula of B\"{u}ttiker Pr\'{e}tre
and Thomas (BPT) \cite{BPT}.
Using the notations of the present Paper
the BPT formula for the current
that comes out of (say) the right lead
can be written as:
\begin{eqnarray}
\fullG^{3j} = \frac{e}{2\pi i}
\trc\left(P\frac{\partial S}{\partial x_j}
S^{\dag}\right)
\end{eqnarray}
where $P$ is the projector on the right lead channels.
For $G^{33}$ the above reduces to the Landauer formula.
The details regarding the relation between
the Kubo formula and the BPT formula will be
published in a separate paper \cite{pmo}. Here we just note that
the derivation is based on a generalization
of the Fisher-Lee approach \cite{datta,fisher,stone}.

Finally it is important to remember that the theory of
driven systems is the corner stone for the analysis
of interaction between ``slow" and ``fast" degrees of freedom.
Assume that that the $x_j$ are in fact dynamical variable,
and that the conjugate momenta are $p_j$.
The standard textbook example is the study
of diatomic molecules. In such case $x_j$ are the locations of the nuclei.
The total Hamiltonian is assumed to be of the general form
\begin{eqnarray}
{\cal H}_{\tbox{total}}=\frac{1}{2M}\sum_j p_j^2 \ + \  {\cal H}(\bm{x})
\end{eqnarray}
where ${\cal H}$ is the Hamiltonian of the
"fast" degrees of freedom (in the context
of molecular physics these are the electrons).
Rather than using the standard basis,
one can use the Born-Oppenheimer basis
$|x,n(\bm{x})\rangle = |x\rangle \otimes |n(\bm{x})\rangle$.
Then the Hamiltonian can be written as
\begin{eqnarray} \nonumber
{\cal H}_{\tbox{total}}=
\frac{1}{2M}\sum_j(p_j-\mbf{A}^j_{nm}(\bm{x}))^2
\ + \  \delta_{nm}E_n(\bm{x})
\end{eqnarray}
where the interaction term is consistent with Eq.(\ref{e_23}).
Thus it is evident that the theory of driven systems
is a special limit of this problem,
which is obtained if we treat the $x_j$ as classical variables.


\ \\ \ \\ 
{\bf Acknowledgments:}
It is my pleasure to thank Yshai Avishai (Ben-Gurion University), Yosi Avron (Technion),
Thomas Dittrich (Colombia), Shmuel Fishman (Technion), Tsampikos Kottos (Gottingen),
and Holger Schantz (Gottingen) for useful discussions.
This research was supported by the Israel Science Foundation (grant No.11/02),
and by a grant from the GIF, the German-Israeli Foundation for Scientific
Research and Development.

\appendix

\clearpage
\section{The Kubo formula: Standard Derivation}

In this Appendix we present an elementary textbook-style
derivation of the Kubo formula. For notational simplicity
we write the Hamiltonian as \mbox{${\cal H}=\bmsf{H}_0-f(t)\bmsf{V}$}.
It is assumed that the system, in the absence
of driving, is prepared in a stationary state $\rho_0$.
In the presence of driving we look for a first order solution
$\rho(t) = \rho_0 +\tilde{\rho}(t)$.
The equation for $\tilde{\rho}(t)$ is:
\begin{eqnarray}
\frac{\partial \tilde{\rho}(t)}{\partial t} \approx
-i[\bmsf{H}_0, \tilde{\rho}(t)] + if(t)[\bmsf{V},\rho_0]
\end{eqnarray}
This equation can be re-written as
\begin{eqnarray} \nonumber
\frac{\partial}{\partial t}
(\bmsf{U}_0(t)^{-1} \tilde{\rho}(t) \bmsf{U}_0(t))
\approx
if(t)[  \bmsf{U}_0(t)^{-1}\bmsf{V}\bmsf{U}_0(t),\rho_0]
\end{eqnarray}
where $\bmsf{U}_0(t)$ is the evolution operator
which is generated by $\bmsf{H}_0$.
The solution of the latter equation is
\begin{eqnarray}
\tilde{\rho}(t)
\ \approx \
\int^{t}   i [\bmsf{V}(-(t{-}t')), \rho_0]  \ f(t')dt'
\end{eqnarray}
where we use the usual definition
of the ``interaction picture" operator
$\bmsf{V}(\tau)=\bmsf{U}_0(\tau)^{-1}\bmsf{V}\bmsf{U}_0(\tau)$.

Consider now the time dependence of the expectation value
\mbox{$\langle \bmsf{F} \rangle_t = \trc(\bmsf{F}\rho(t))$}
of an observable. Disregarding the zero order contribution,
the first order expression is
\begin{eqnarray} \nonumber
\langle \bmsf{F} \rangle_t
\ &\approx& \
\int^{t}  i \trc\left(\bmsf{F} [\bmsf{V}(-(t{-}t')), \rho_0]\right)  \ f(t')dt'
\\ \nonumber
&=& \
\int^{t}   \alpha(t-t')  \ f(t')dt'
\end{eqnarray}
where the response kernel $\alpha(\tau)$ is
defined for $\tau>0$ as
\begin{eqnarray}
\alpha(\tau) &=& i\ \trc\left(\bmsf{F} [\bmsf{V}(-\tau), \rho_0]\right)
\nonumber \\
\ &=& i\ \trc\left([\bmsf{F}, \bmsf{V}(-\tau)] \rho_0\right)
\nonumber \\
\ &=& i \langle [\bmsf{F}, \bmsf{V}(-\tau)] \rangle
\nonumber \\
\ &=& i \langle [\bmsf{F}(\tau), \bmsf{V}] \rangle
\end{eqnarray}
We have used above the cyclic property of the trace operation;
the stationarity $\bmsf{U}_0\rho_0\bmsf{U}_0^{-1}=\rho_0$ of
the unperturbed state; and the definition
$\bmsf{F}(\tau)=\bmsf{U}_0(\tau)^{-1}\bmsf{F}\bmsf{U}_0(\tau)$.

\newpage
\section{Remarks regarding the generalized susceptibility}

In this appendix we would like to further illuminate the relation
between the generalized susceptibility and the conductance matrix.
The generalized susceptibility $\chi^{kj}(\omega)$ is
the Fourier transform of the causal response kernel $\alpha^{kj}(\tau)$.
Therefore it is an analytic function in the upper
half of the complex $\omega$ plan, whose real and imaginary
parts are related by Hilbert transforms (Kramers-Kronig relations):
\begin{eqnarray}
\chi_0^{kj}(\omega)  \equiv  \re [\chi^{kj}(\omega)] =
\int_{-\infty}^{\infty} \frac{\im[\chi^{kj}(\omega')]}{\omega'-\omega} \ \frac{d\omega'}{\pi}
\end{eqnarray}
The imaginary part of $\chi^{kj}(\omega)$
is the sine transforms of $\alpha^{kj}(\tau)$,
and therefore it is proportional to $\omega$
for small frequencies.
Consequently it is convenient to write the
Fourier transformed version of Eq.(\ref{e_3}) as
\begin{eqnarray} \label{e3}
[\langle F^k \rangle]_{\omega} \ = \ \sum_j
\chi_0^{kj}(\omega) [x_j]_{\omega}
-\mu^{kj}(\omega) [\dot{x}_j]_{\omega}
\end{eqnarray}
where the dissipation coefficient is defined as
\begin{eqnarray}
\mu^{kj}(\omega)  =
\frac{\im[\chi^{kj}(\omega)]}{\omega} =
\int_0^{\infty} \alpha^{kj}(\tau) \frac{\sin(\omega\tau)}{\omega}d\tau
\end{eqnarray}
In this paper we ignore the first term in Eq.(\ref{e3})
which signify the non-dissipative in-phase response.
Rather we put the emphasis on the ``DC limit" ($\omega\rightarrow 0$)
of the second term. Thus the conductance
matrix \mbox{$\fullG^{kj}=\mu^{kj}(\omega \rightarrow 0)$}
is just a synonym for the term ``dissipation coefficient".
However, ``conductance" is a better (less misleading) terminology:
it does not have the (wrong) connotation
of being  specifically  associated with dissipation,
and consequently it is less confusing to say that
it contains a (non-dissipative) adiabatic component.

For systems where time reversal symmetry is broken
due to the presence of a magnetic field ${\cal B}$,
the response kernel, and consequently the generalized susceptibility
and the conductance matrix satisfies the Onsager reciprocity relations
\begin{eqnarray}
\alpha^{ij}(\tau, -{\cal B}) \ &=& \ [\pm] \ \alpha^{ji}(\tau, {\cal B}) \\
\chi^{ij}(\omega, -{\cal B}) \ &=& \ [\pm] \ \chi^{ji}(\omega, {\cal B}) \\
\fullG^{ij} (-{\cal B}) \  &=& \    [\pm]   \ \fullG^{ji}({\cal B})
\end{eqnarray}
where the plus (minus)  applies if the signs of $F^i$ and $F^j$
transform (not) in the same way under time reversal.
These reciprocity relations follow from
the Kubo formula (Eq.(\ref{e_kubo}),
using
$K^{ij}(-\tau,-{\cal B}) = - [\pm] K^{ij}(\tau,{\cal B}) $,
together with the trivial identity
$K^{ij}(-\tau,{\cal B}) = - K^{ji}(\tau,{\cal B}) $.
In Section.~9 we discuss the implications of the reciprocity relations
in the context of pumping.

\clearpage
\onecolumngrid
\section{Expressions for $\bmsf{B}$ and $\fullG$}

The functions $C^{ij}(\tau)$  and $K^{ij}(\tau)$ are
the expectation values of hermitian operators.
Therefore they are real functions. It follows that the
real part of their Fourier transform
is a symmetric function with respect to $\omega$,
while the imaginary part of their Fourier transform
is anti symmetric with respect to $\omega$.
By definition they satisfy $C^{ij}(\tau)  = C^{ji}(-\tau)$
and $K^{ij}(\tau)  = -K^{ji}(-\tau) $.
It is convenient to regard them as the real and imaginary parts
of one complex function $\Phi^{ij}(\tau)$. Namely,
\begin{eqnarray}
\Phi^{ij}(\tau) &=& \langle F^i(\tau)F^j(0)\rangle \ \ = \ \ C^{ij}(\tau)-i\frac{\hbar}{2}K^{ij}(\tau) \\
C^{ij}(\tau) &=& \frac{1}{2}\left(\Phi^{ij}(\tau) + \Phi^{ji}(-\tau)\right) \\
K^{ij}(\tau) &=& \frac{i}{\hbar} \left(\Phi^{ij}(\tau) - \Phi^{ji}(-\tau)\right)
\end{eqnarray}

It is possible to express the decomposition $\fullG^{ij}=\symG^{ij}+\mbf{B}^{ij} $
in terms of $\tilde{K}^{ij}(\omega)$. Using the definition Eq.(\ref{e_33}) we get:
\begin{eqnarray} \label{eB4}
\fullG^{ij}
\ \ = \ \
\int_0^{\infty}K^{ij}(\tau)\tau d\tau
\ \ = \ \
{-}\int_{-\infty}^{\infty}\frac{\re[\tilde{K}^{ij}(\omega)]}{\omega^2}
\frac{d\omega}{2\pi} +
\left[\frac{1}{2}\frac{\im[\tilde{K}^{ij}(\omega)]}{\omega}\right]_{\omega{=}0}
\end{eqnarray}
The first term is antisymmetric with respect to its indexes,
and is identified as $\mbf{B}^{ij}$.
The second term is symmetric with respect to its indexes,
and is identified as $\symG^{ij}$.
The last step in the above derivation involves
the following identity that hold for any real function $f(\tau)$
\begin{eqnarray}
\int_0^{\infty}f(\tau)\tau d\tau
=
\int_{-\infty}^{\infty}\frac{d\omega}{2\pi}\tilde{f}(\omega)
\int_0^{\infty}\eexp{-i\omega\tau}\tau d\tau
=
\int_{-\infty}^{\infty}\frac{d\omega}{2\pi}
\tilde{f}(\omega)
\left(-\frac{1}{\omega^2}+i\pi\delta'(\omega)\right)
=
\nonumber \\
=
\int_{-\infty}^{\infty}\frac{d\omega}{2\pi}
\left(-\frac{\re[\tilde{f}(\omega)]}{\omega^2}
-\pi\im[\tilde{f}(\omega)]\delta'(\omega)\right)
=
{-}\int_{-\infty}^{\infty}\frac{\re[\tilde{f}(\omega)]}{\omega^2}
\frac{d\omega}{2\pi} +
\left[\frac{1}{2}\frac{\im[\tilde{f}(\omega)]}{\omega}\right]_{\omega{=}0}
\end{eqnarray}
Note that $\im[\tilde{f}(\omega)]$ is the sine transform of $f(\tau)$,
and therefore it is proportional to $\omega$ is the limit of small frequencies.

It is of practical value to re-derive Eq.(\ref{eB4})
by writing $\Phi^{ij}(\tau)$ using the energies $E_n$
and  the matrix elements~$F^i_{nm}$.
Then we can get from it straightforwardly (using the definitions)
all the other expressions. Namely:
\begin{eqnarray}
\Phi^{ij}(\tau) \ &=& \
\sum_n f(E_n) \sum_{m} F^i_{nm}F^j_{mn}
\exp\left(-i\frac{E_m{-}E_n}{\hbar}t \right)
\\
\tilde{\Phi}^{ij}(\omega)  \ &=& \
\sum_n f(E_n) \sum_{m} F^i_{nm}F^j_{mn}
\ 2\pi\delta\left(\omega-\frac{E_m{-}E_n}{\hbar} \right)
\\
\chi^{ij}(\omega) \ &=& \ \sum_{n,m} f(E_n)\left(
\frac{-F^i_{nm}F^j_{mn}}{\hbar\omega{-}(E_m{-}E_n){+}i0}
+\frac{F^j_{nm}F^i_{mn}}{\hbar\omega{+}(E_m{-}E_n){+}i0}\right)
\\
\symG^{ij} \ &=& \ -2\pi\hbar  \sum_n f(E_n) \sum_{m(\ne n)}
\re\left[F^i_{nm}F^j_{mn}\right]
\ \delta'(E_m-E_n)
\\ \label{eB10}
\mbf{B}^{ij} \ &=& \ 2\hbar  \sum_n f(E_n) \sum_{m(\ne n)}
\frac{\im\left[
F^i_{nm}F^j_{mn}\right]}
{(E_m-E_n)^2}
\end{eqnarray}
One observes that the expression for $\mbf{B}^{ij}$ coincides
with the adiabatic transport result Eq.(\ref{e_30}).
Alternatively this identification can be obtained by expressing
the sum in Eq.(\ref{eB10}) as an integral, getting form it
the first term in Eq.(\ref{eB4}):
\begin{eqnarray}
\mbf{B}^{ij} \ \  = \ \
\frac{2}{\hbar}\int_{-\infty}^{\infty}
\frac{\im[\tilde{\Phi}^{ij}(\omega)]}{\omega^2}\frac{d\omega}{2\pi}
\ \ = \ \
\frac{2}{\hbar}\int_{-\infty}^{\infty}
\frac{\im[\tilde{C}^{ij}(\omega)]-\frac{\hbar}{2}\re[ \tilde{K}^{ij}(\omega) ]}
{\omega^2}\frac{d\omega}{2\pi}
\ \ = \ \
-\int_{-\infty}^{\infty}
\frac{\re[\tilde{K}^{ij}(\omega)]}{\omega^2}\frac{d\omega}{2\pi}
\end{eqnarray}

\clearpage
\onecolumngrid
\section{Expressing $\tilde{K}(\omega)$ using  $\tilde{C}(\omega)$}

We can use the following manipulation in order
to relate  $\tilde{K}^{ij}(\omega)$ to $\tilde{C}^{ij}(\omega)$,
\begin{eqnarray} \label{eC1}
\tilde{K}^{ij}(\omega)&=& \sum_n f(E_n) \ \tilde{K}^{ij}_n(\omega)
\\ \nonumber &=&
\frac{i}{\hbar} 2\pi \sum_{nm}f(E_n)
(F^i_{nm}F^j_{mn}\delta(\omega+\omega_{nm})-F^j_{nm}F^i_{mn}\delta(\omega-\omega_{nm}))
\\ \nonumber  &=&
\frac{i}{\hbar} 2\pi \sum_{nm}f(E_m)
(-F^i_{nm}F^j_{mn}\delta(\omega+\omega_{nm})+F^j_{nm}F^i_{mn}\delta(\omega-\omega_{nm}))
\\ \nonumber  &=&
\frac{i}{\hbar} 2\pi \sum_{nm} \frac{f(E_n)-f(E_m)}{2}
(F^i_{nm}F^j_{mn}\delta(\omega+\omega_{nm})-F^j_{nm}F^i_{mn}\delta(\omega-\omega_{nm}))
\\ \nonumber  &=&
-i \omega \pi \sum_{nm}\frac{f(E_n)-f(E_m)}{E_n-E_m}
(F^i_{nm}F^j_{mn}\delta(\omega+\omega_{nm})+F^j_{nm}F^i_{mn}\delta(\omega-\omega_{nm}))
\\ \nonumber &=&
-i\omega\sum_{n} f'(E_n) \ C^{ij}_n(\omega)
\end{eqnarray}
where we use the notation $\omega_{nm}=(E_n-E_m)/\hbar$.
The third line differs from the second line by permutation of the dummy
summation indexes, while the fourth line is the sum of the second
and the third lines divided by 2. In the last equality we assume small $\omega$.
If the levels are very dense, then we can replace the summation by integration,
leading to the relation:
\begin{eqnarray}
\int g(E)dE \ f(E) \ \tilde{K}^{ij}_{E}(\omega)
\ \ = \ \
-i \omega \int g(E)dE \ f'(E) \ \tilde{C}^{ij}_{E}(\omega)
\end{eqnarray}
where $\tilde{K}^{ij}_{E}(\omega)$ and $\tilde{C}^{ij}_{E}(\omega)$
are microcanonically smoothed functions.
Since this equality hold for any smoothed $f(E)$, it follows
that the following relation holds (in the limit $\omega\rightarrow0$):
\begin{eqnarray}
\tilde{K}^{ij}_E(\omega)
\  = \
i\omega \frac{1}{g(E)}\frac{d}{dE}\left[g(E)C^{ij}_E(\omega)\right]
\end{eqnarray}
If we do not assume small $\omega$, but instead assume canonical state,
then a variation on the last steps in Eq.(\ref{eC1}),
using the fact that $(f(E_n){-}f(E_m))/ (f(E_n){+}f(E_m))= \tanh((E_n{-}E_m)/(2T))$
is an odd function, leads to the relation
\begin{eqnarray}
\tilde{K}^{ij}_T(\omega)  \ = \  i\omega \times
\frac{1}{\hbar\omega}\tanh\left(\frac{\hbar\omega}{2T}\right)  \  C^{ij}_T(\omega)
\end{eqnarray}
Upon substitution of the above expressions in the Kubo formula for $\symG^{ij}$,
one obtains the Fluctuation-Dissipation relation.


\ \\ \ \\
\line(1,0){490}
\ \\ \ \\

\twocolumngrid
\section{The Kubo formula and the diffusion in energy space}

The illuminating derivation of Eq.(\ref{e43}) is based on
the observation that energy absorption is related to
having diffusion in energy space. Let us assume that
the probability distribution $\rho(E)=g(E)f(E)$ of the energy
satisfies the following diffusion equation:
\begin{eqnarray} \label{eD1}
\frac{\partial \rho}{\partial t} \ = \
\frac{\partial}{\partial E}
\left(g(E)D_E \frac{\partial}{\partial E}
\left(\frac{1}{g(E)}\rho\right)\right)
\end{eqnarray}
The energy of the system is
$\langle {\cal H} \rangle=\int E \rho(E)dE$.
It follows that the rate of energy absorption is
\begin{eqnarray}
\frac{d}{dt}\langle {\cal H} \rangle
= - \int_0^{\infty} dE \ g(E) \ D_E
\ \frac{\partial}{\partial E}
\left(\frac{\rho(E)}{g(E)}\right)
\end{eqnarray}
For a microcanonical preparation we get
\begin{eqnarray} \label{e48}
\frac{d}{dt}\langle {\cal H} \rangle
\ = \
\frac{1}{g(E)}
\frac{d}{dE}\left[g(E) \ D_E \right]
\end{eqnarray}
This diffusion-dissipation relation reduces
immediately to the fluctuation-dissipation relation
if we assume that the diffusion in energy
space due to the driving is given by
\begin{eqnarray} \label{e49}
D_E \ = \
\frac{1}{2}\sum_{ij}
\tilde{C}^{ij}_E(\omega{\rightarrow}0)
\ \dot{x}_i \dot{x}_j
\end{eqnarray}
Thus it is clear that a theory for linear response
should establish that there is a diffusion process in energy
space due to the driving, and that the diffusion
coefficient is given by Eq.(\ref{e49}).
More importantly, this approach also allows
treating cases where the expression
for $D_E$ is non-perturbative,
while the diffusion-dissipation relation Eq.(\ref{e48}) still holds!

A full exposition (and further  reference) for this route
of derivation can be found in \cite{vrn,dsp,crs,frc}.
Here we shall give just the classical derivation of Eq.(\ref{e49}),
which is extremely simple. We start with the identity
\begin{eqnarray}
\frac{d}{dt}\langle {\cal H} \rangle
\ = \ \left\langle  \frac{\partial {\cal H}}{\partial t} \right\rangle \ =
\ -\sum_k    \dot{x}_k   F^k(t)
\end{eqnarray}
Assuming (for presentation purpose) that
the rates $\dot{x}_k$ are constant numbers,
it follows that energy changes are related
to the fluctuating $F^k(t)$ as follows:
\begin{eqnarray}
\delta E \ = \ \langle {\cal H} \rangle_t - \langle {\cal H} \rangle_0 \ =
\ -\sum_k    \dot{x}_k \int_0^t F^k(t')dt'
\end{eqnarray}
Squaring this expression, and performing
microcanonical averaging over initial conditions we obtain:
\begin{eqnarray}
\delta E^2(t) \ = \  \sum_{ij} \dot{x}_i\dot{x}_j \int_0^t\int_0^t
C^{ij}_E(t''-t') dt' dt''
\end{eqnarray}
where $C^{ij}_E(t''-t')= \langle F^i(t')  F^j(t'') \rangle$ is the
correlation function. For very short times this equation
implies ``ballistic'' spreading ($\delta E^2 \propto t^2$)
while on {\em intermediate time scales} it leads
to diffusive spreading \mbox{$\delta E^2(t) =2D_E t$}, where
\begin{eqnarray}
D_E \ = \ \frac{1}{2} \sum_{ij} \dot{x}_i\dot{x}_j \int_{-\infty}^{\infty}
C^{ij}_E(\tau) d\tau
\end{eqnarray}
The latter result assumes a short correlation time.
This is also the reason that the integration
over $\tau$ can be extended form
$-\infty$ to $+\infty$. Hence we get Eq.(\ref{e49}).
We note that for long times the systems deviates
significantly from the initial microcanonical preparation.
Hence, for long times, one should justify the
use of the diffusion equation (\ref{eD1}).
This leads to the classical slowness condition
which is discussed in Ref.\cite{frc}.



\clearpage
\onecolumngrid

\centerline{\epsfig{figure=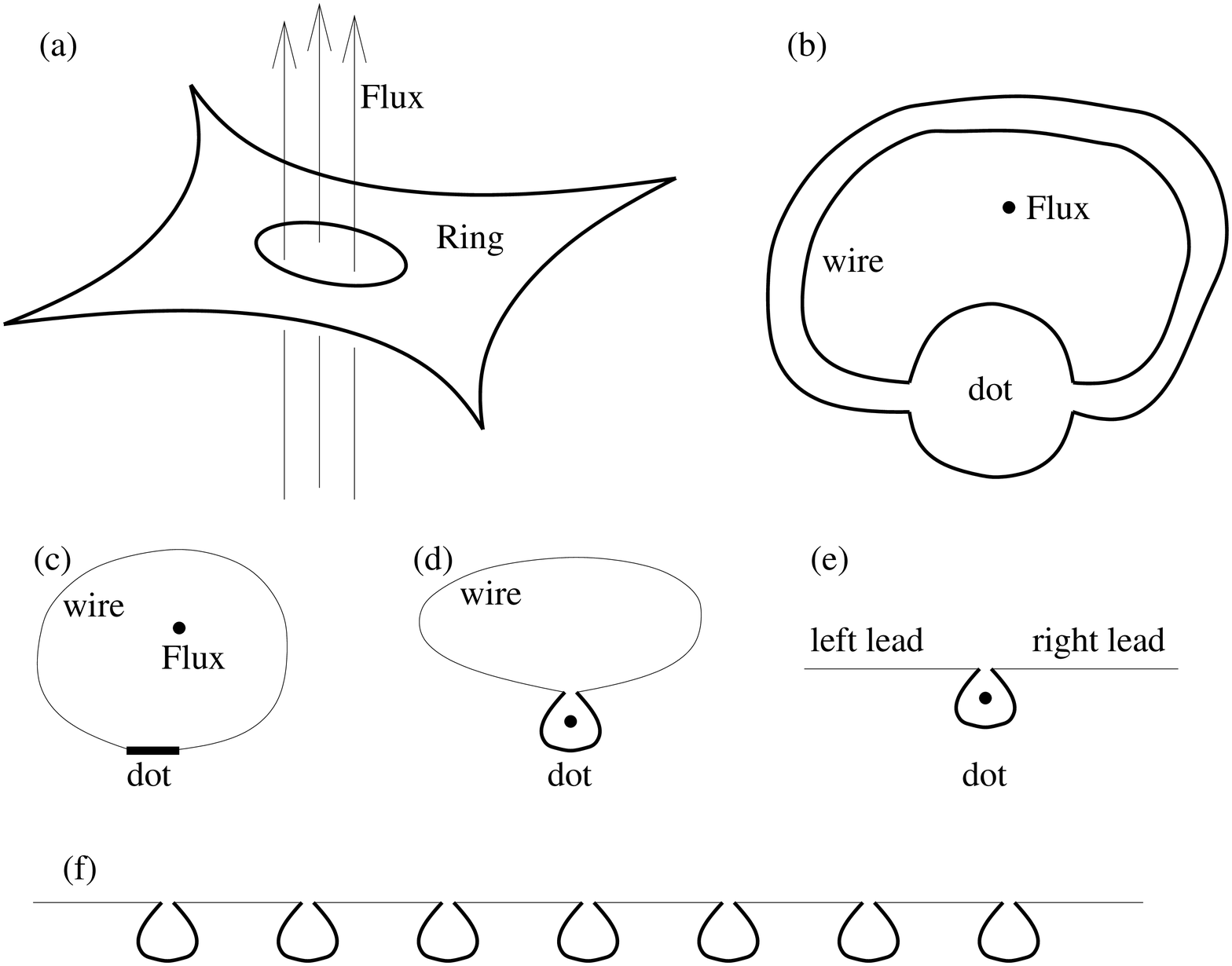,width=0.55\hsize}}
{\footnotesize FIG1.
Illustration of a ring system (a).
The shape of the ring is controlled
by some parameters $x_1$ and $x_2$.
The flux through the ring is $x_3=\Phi$.
A system with equivalent topology,
and abstraction of the model are
presented in (b) and (c).
The ``dot" can be represented by an $S$ matrix
that depends on $x_1$ and $x_2$. In (d) also the
flux $x_3$ is regarded as a parameter of the dot.
If we cut the wire in (d) we get the open
two lead geometry of (e). If we put many such
units in series we get the period system in (f).
}

\ \\

\centerline{
\epsfig{figure=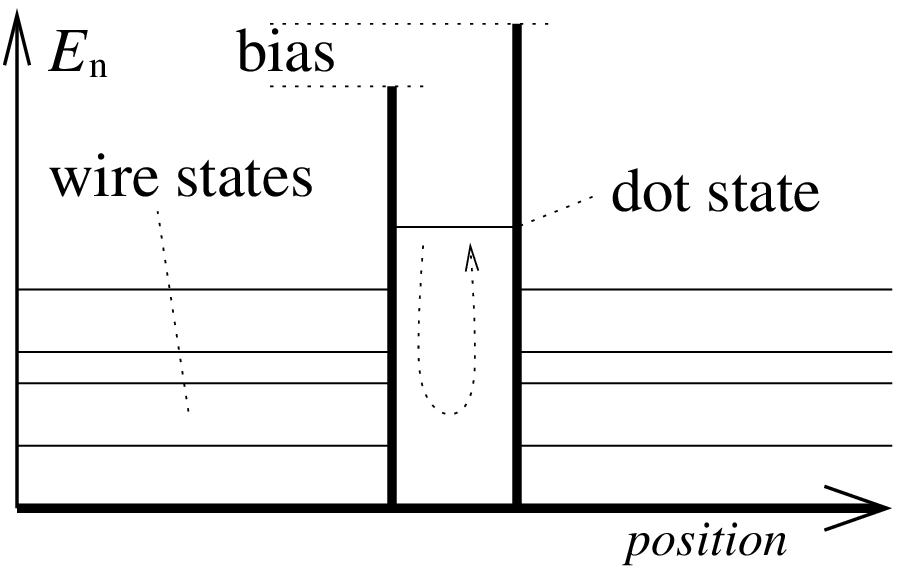,height=0.2\hsize}
\epsfig{figure=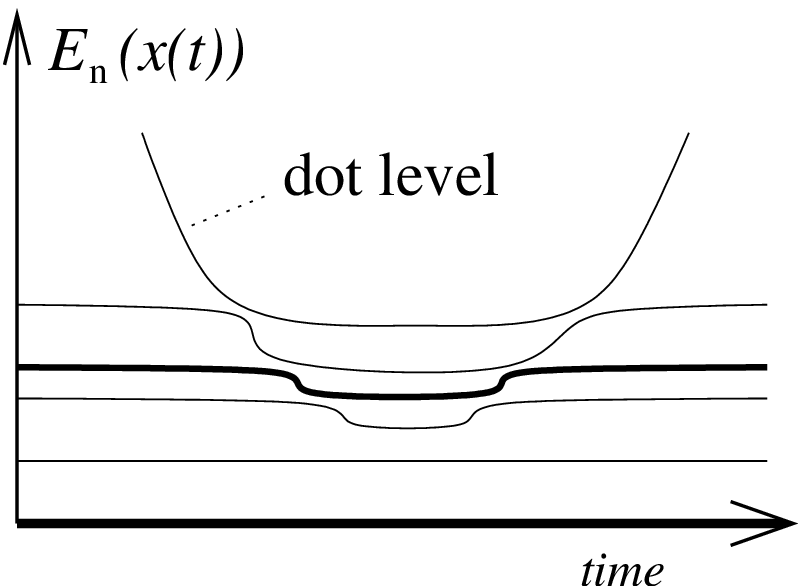,height=0.2\hsize}
}
{\footnotesize FIG2.
Schematic illustration of quantum pumping
in a closed wire-dot system. The net charge via the third
level (thick solid line on the right) is vanishingly
small: As the dot potential is lowered an electron
is taken from the left side (first avoided crossing),
and then emitted back to the left side
(second avoided crossing).
Assuming that the bias is inverted before the
dot potential is raised back, only the second level
carry a net charge $Q={\cal O}(1)$.}

\ \\

\centerline{\epsfig{figure=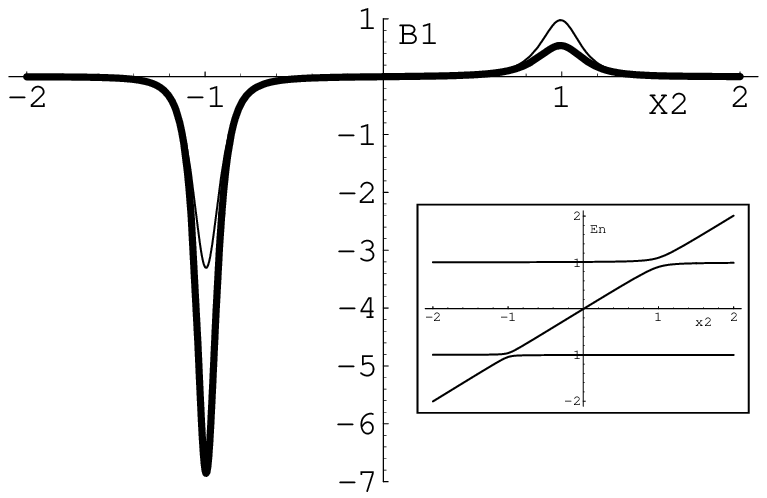,width=0.5\hsize}}
\
{\footnotesize FIG3.
The first component of the $\vec{\bm{B}}$ field
for a particle in the middle level of
the 3~site lattice model. It is plotted
as a function of the dot potential $x_2=u$.
The other parameters are $\phi=0$, and $c_1=0.1$,
while $c_2=0.04$ for the thick line and
$c_2=0.02$ for the thin line.
In the limit $c_2 \rightarrow 0$,
all the charge that is transfered from the
left side into the dot during the first avoided crossing,
is emitted back into the left side during the second
avoided crossing.
Inset: The eigenenergies $E_n(\bm{x})$ for
the $c_2=0.04$ calculation.}


\begin{thebibliography}{99}



\bibitem{landau}
L.D. Landau and E.M. Lifshitz,
{\em Statistical physics}, (Buttrworth Heinemann 2000).

\bibitem{imry}
Y. Imry, {\em Introduction to Mesoscopic Physics}
(Oxford Univ. Press 1997), and references therein.

\bibitem{datta}
S. Datta, {\em Electronic Transport in Mesoscopic Systems}
(Cambridge University Press 1995).

\bibitem{marcus_rev}
L.P. Kouwenhoven et al,
Proc. of Advanced Study Inst. on Mesoscopic
Electron Transport, edited by L.L. Sohn,
L.P. Kouwenhoven and G. Schon (Kluwer 1997).



\bibitem{thouless}
D. J. Thouless, Phys. Rev. {\bf B27}, 6083 (1983).

\bibitem{BPT}
M. B\"{u}ttiker et al, Z. Phys. {\bf B94}, 133 (1994). \
P. W. Brouwer, Phys. Rev. {\bf B58}, R10135 (1998). \
J. E. Avron et al, Phys. Rev. B {\bf 62}, R10 618 (2000).

\bibitem{aleiner}
T. A. Shutenko, I. L. Aleiner and B. L. Altshuler, Phys. Rev. {\bf B61}, 10366 (2000).

\bibitem{barriers}
Y. Levinson, O. Entin-Wohlman, and P. Wolfle, cond-mat/0010494. \
M. Blaauboer and E.J. Heller, Phys. Rev. B 64, 241301(R) (2001).

\bibitem{ora}
O. Entin-Wohlman, A. Aharony, and Y. Levinson Phys. Rev. B 65, 195411 (2002).

\bibitem{pmo}
D. Cohen, cond-mat/0304678.

\bibitem{ratch} P.Reimann, Phys. Rep. {\bf 361}, 57 (2002). \
Special issue, Appl. Phys. A {\bf 75} (2002). \
P.Reimann, M. Grifoni, and P.Hanggi, Phys. Rev. Lett. {\bf 79}, 10 (1997).

\bibitem{ratchets}
H. Schanz, M.F. Otto, R. Ketzmerick, and T. Dittrich, Phys. Rev. Lett. {\bf 87}, 070601 (2001).




\bibitem{berry}
M.V. Berry, Proc. R. Soc. Lond. A {\bf 392}, 45 (1984).


\bibitem{avron}
J. E. Avron and L. Sadun, Phys. Rev. Lett. {\bf 62}, 3082 (1989). \
J. E. Avron and L. Sadun,  Ann. Phys. {\bf 206}, 440 (1991). \
J.E. Avron, A. Raveh, and B. Zur  Rev. Mod. Phys. {\bf 60}, 873  (1988).

\bibitem{robbins}
J.M. Robbins and M.V. Berry, J. Phys. A {\bf 25}, L961 (1992). \
M.V. Berry and J.M. Robbins, Proc. R. Soc. Lond. A {\bf 442}, 659 (1993). \
M.V. Berry and E.C. Sinclair, J. Phys. A {\bf 30}, 2853 (1997).




\bibitem{vrn}
D. Cohen in {\em New directions in quantum chaos},
Proceedings of the International School of Physics ``Enrico Fermi", Course CXLIII,
Edited by G. Casati, I. Guarneri and U. Smilansky,
(IOS Press, Amsterdam 2000).

\bibitem{dsp}
D. Cohen in {\em Dynamics of Dissipation},
Proceedings of the 38th Karpacz Winter School of Theoretical Physics,
Edited by P. Garbaczewski and R. Olkiewicz,
(Springer, 2002)

\bibitem{crs}
D. Cohen, Phys. Rev. Lett. {\bf 82}, 4951 (1999). \
D. Cohen and T. Kottos, Phys. Rev. Lett. 85, 4839 (2000).

\bibitem{frc}
D. Cohen, Annals of Physics 283, 175 (2000).

\bibitem{wilk}
M. Wilkinson, J. Phys. A {\bf 21}, 4021 (1988). \
M. Wilkinson and E.J. Austin, J. Phys. A {\bf 28}, 2277 (1995).




\bibitem{ophir}
O.M. Auslaender and S. Fishman, Phys. Rev. Lett. {\bf 84}, 1886 (2000). \
O.M. Auslaender and S. Fishman,  J. Phys. A {\bf 33}, 1957 (2000).

\bibitem{qkr}
S. Fishman in {em Quantum Chaos},
Proceedings of the International School
of Physics "Enrico Fermi", Course CXIX,
Ed. G. Casati, I. Guarneri and U. Smilansky
(North Holland 1991).

\bibitem{dil}
A. Barnett, D. Cohen and E.J. Heller, Phys. Rev. Lett. {\bf 85}, 1412 (2000).

\bibitem{wlf}
A. Barnett, D. Cohen and E.J. Heller, J. Phys. A {\bf 34}, 413 (2001).

\bibitem{fisher}
D.S. Fisher and P.A. Lee, Phys. Rev. B {\bf 23}, 6851 (1981).

\bibitem{stone}
H.U. Baranger and A.D. Stone, Phys. Rev. B {\bf 40}, 8169 (1989).

\end{thebibliography}
\end{document}